\documentclass[reprint,amsmath,amssymb,aps,prd]{revtex4-2}

\usepackage[utf8]{inputenc}
\usepackage[title]{appendix}
\usepackage{graphicx}
\usepackage{amssymb}
\usepackage{amsmath}
\usepackage{subfigure}
\usepackage{bm}
\usepackage{booktabs}
\usepackage{xcolor}
\usepackage{adjustbox}
\usepackage{hyperref}
\usepackage{xcolor}
\begin{document}
\title{Feasibility Study of Pion and Kaon Structure via the Sullivan Process at EicC}
\author{Zongyang Lu$^{a}$}
\author{Zihan Yu$^{a}$}
\author{Ting Lin$^{a}$}
\author{Yu-Tie Liang$^{b, c}$}
\author{Rong Wang$^{b,c}$}
\author{Wan Chang$^{d}$}
\author{Weizhi Xiong$^{a}$}
\email{xiongw@sdu.edu.cn}

\affiliation{$^a$Key Laboratory of Particle Physics and Particle Irradiation (MOE),
Institute of Frontier and Interdisciplinary Science,Shandong University, Qingdao, Shandong 266237, China}

\affiliation{$^b$Institute of Modern Physics, Chinese Academy of Sciences, Lanzhou 730000, China}

\affiliation{$^c$University of Chinese Academy of Sciences, Beijing 100049, China}
\affiliation{$^d$School of Physics and Electronic Engineering, Nanyang Normal University, Nanyang 473061, China}

\begin{abstract}
The Electron--Ion Collider in China (EicC) provides an excellent opportunity to explore the internal structure of pions and kaons via the Sullivan process in deep-inelastic scattering (DIS). In this study, we present detailed projections for the pion and kaon structure functions, $F_2^\pi$ and $F_2^K$, at EicC, with a focus on both statistical and systematic uncertainties. Leveraging EicC's high luminosity and broad kinematic coverage, the accessible kinematic region is extended beyond previous measurements. The projected statistical uncertainties for $F_2^\pi$ and $F_2^K$ are below 5\% and 8\%, respectively, across most kinematic bins. Systematic uncertainties arising from detector effects have been carefully evaluated. These results significantly enhance the precision of meson structure function measurements and provide important constraints on theoretical models of meson parton distributions. Moreover, this study bridges the gap between fixed-target and collider-era measurements, highlighting the pivotal role of EicC in advancing our understanding of hadronic structure.

\end{abstract}

\maketitle
\newpage
\quad

\section{Introduction}

The pion is the lightest meson, while the kaon is the lightest meson containing a strange quark in the Standard Model, with each composed of a quark–antiquark pair bound by the strong interaction. Their simplicity makes them ideal probes of Quantum Chromodynamics (QCD), the theory describing the strong force. Pions are nearly ideal realizations of pseudo-Nambu–Goldstone bosons~\cite{Nambu:1960tm,Goldstone:1961eq}, arising from the spontaneous breaking of chiral symmetry in QCD, while kaons provide an essential window into the interplay between light and strange quarks. Together, they form a unique platform for exploring the low-energy domain of QCD, where perturbative methods break down and nonperturbative dynamics dominate~\cite{Deur:2016cxb, Deur:2016tte,deTeramond:2016htp,Roberts:2021nhw}.\par

Recent progress in lattice QCD~\cite{Gao:2020ito,Gao:2021xsm,Ding:2024lfj,Chen:2019lcm,Lin:2020ssv}, Dyson–Schwinger method~\cite{Roberts:1994dr,Maris:1997hd,Chang:2013pq,Shi:2018mcb, Bednar:2018mtf}, and effective field theories~\cite{Machleidt:2011zz,Weinberg:1991um, Hutauruk:2018zfk}, together with high-energy scattering experiments \cite{CLAS:2012qga,HERMES:2007hrc,Dally:1981ur,Dally:1982zk,Amendolia:1984nz,NA7:1986vav,Favart:2015umi,Brown:1973wr,Arnold:1974ik,Bebek:1977pe,Ackermann:1977rp,Brauel:1979zk,JeffersonLabFpi:2007vir,JeffersonLabFpi:2007vir,JeffersonLabFpi-2:2006ysh,Carmignotto:2018uqj}, has enabled increasingly precise investigations of pion and kaon structure. These studies have advanced our knowledge of their electromagnetic form factors, parton distribution functions (PDFs), and generalized parton distributions (GPDs). Data from facilities such as Jefferson Lab~\cite{Accardi:2023chb,Horn:2007ug,JeffersonLab:2008jve,TDIS,TDIS-kaon}, 
HERA \cite{DeBoer:2016tnn} experiments including H1 \cite{H1:2010hym} and ZEUS~\cite{ZEUS:2002gig}, 
CERN fixed-target experiments~\cite{Quintans:2022utc,Adams:2018pwt}, 
and the future Electron–Ion Collider (EIC)~\cite{Aguilar:2019teb,Arrington:2021biu,AbdulKhalek:2021gbh,Aguilar:2019teb}, 
combined with theoretical developments, have revealed that mesons are far from pointlike: 
their internal structure is significantly more intricate than suggested by constituent-quark models~\cite{Roberts:2021nhw,Ding:2022ows}. Despite these advances, a comprehensive understanding of their partonic structure remains incomplete. 

A powerful method to access the structure of these unstable mesons is the Sullivan process~\cite{Gao:2021xsm,Ding:2024lfj,RuizArriola:2024gwb, Sullivan:1971kd}, in which deep-inelastic scattering off the virtual meson cloud of a nucleon, effectively providing a pion or kaon target. This approach overcomes the limitation of mesons’ short lifetimes and has been validated in previous experiments for pions~\cite{H1:2010hym,ZEUS:2002gig,H1:1998vul}.  For kaons, the method presents additional challenges due to the suppressed strange-quark component in the nucleon’s meson cloud, but it remains a viable strategy for extracting their partonic structure~\cite{Roberts:2021nhw}.\par

The EicC~\cite{Anderle:2021wcy} offers a unique opportunity to probe the internal structure of hadrons with particular emphasis on the sea-quark region. In this work, we develop a production model based on the Sullivan process to investigate the structure of pions and kaons. By examining the relevant kinematics and detector requirements, we present detailed projection studies of their structure function at EicC. The remainder of this paper is organized as follows: Section 2 outlines the theoretical framework and the implementation of the Sullivan process in our analysis. Section 3 describes the conceptual design of the experimental setup and the strategies for event selection and analysis. Section 4 presents the main results, including comparisons with theoretical predictions and existing data. Finally, Section 5 summarizes the key findings and discusses prospects for future research.

\section{Kinematics and cross section}

\begin{figure*}[htbp]
    \centering
    \includegraphics{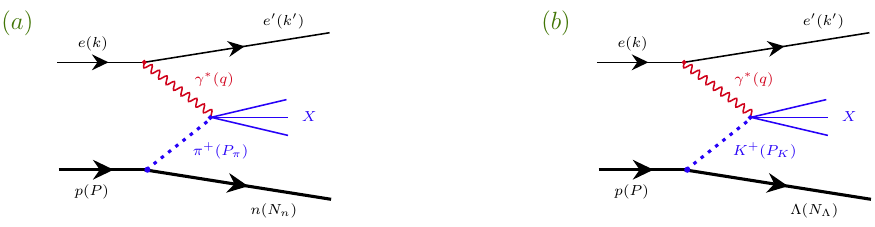}
    \caption{Sullivan process in DIS. $(a)$ Pion case: the proton emits a virtual $\pi^+$ and transitions into a neutron. $(b)$ Kaon case: the proton emits a virtual $K^+$ and transitions into a $\Lambda$.}
    \label{Feynmandiagram}
\end{figure*}

Figure \ref{Feynmandiagram} shows the Feynman diagrams for semi-inclusive leading neutron $(a)$ and $\Lambda$ $(b)$ production in collisions via the Sullivan process. In this framework, the process $ep\to eX\mathcal{B}$, where $\mathcal{B}$ represents the leading baryon, provides access to the structure of unstable mesons. The process is fully described by four kinematic variables:

\begin{equation*} 
Q^2 = -q^2 = -(k-k^\prime)^2,
\end{equation*}  
\begin{equation*}
x_B = \frac{Q^2}{2P\cdot q}, 
\end{equation*} 
\begin{equation*} 
y = \frac{q\cdot P}{k\cdot P} \simeq \frac{Q^2}{s x_B},
\end{equation*}
\begin{equation} 
W^2 = (P+k-k^\prime)^2 = m^2_p + \frac{Q^2(1-x_B)}{x_B}, 
\end{equation}
 
where $k$, $P$, $k^\prime$, and $N$ represent the four-momenta of the incoming electron $(e)$, proton $(p)$, scattered electron $(e^\prime)$, and leading baryon $(n~\text{or}~\Lambda)$, respectively, and $m_p$ is the proton mass. When probing the pion structure, the final-state baryon is a neutron ($\mathcal{B}=n$); for the kaon structure, it is a $\Lambda$ ($\mathcal{B}=\Lambda$). In this process, the energy fraction of the final-state hadron $x_L$, and the squared four-momentum transfer of the target proton $t$, are expressed as:

\begin{equation}
    x_L = \frac{N\cdot k}{P \cdot k},~~~ t = (P-N)^2
\end{equation}
respectively. The exchanged virtual pion or kaon can then be treated as an effective target. Analogous to the Bjorken variable, the momentum fraction of the struck constituent inside the meson is defined as:
\begin{equation}
    x_\mathcal{M}= \frac{Q^2}{2P_\mathcal{M}\cdot q}=\frac{x_B}{1-x_L}
\end{equation}

Thus, the differential cross section for leading neutron or $\Lambda$ production in the process $ep \to e^\prime X \mathcal{B}$ can be written as~\cite{H1:2010hym, ZEUS:2002gig,Holtmann:1995qev}
\begin{widetext}
\begin{align}
\frac{d^4 (\sigma \to e X \mathcal{B})}{dx_B dQ^2 dx_L dt} 
&= \frac{4\pi \alpha^2}{x_B Q^4} \left(1-y+\frac{y^2}{2}\right) F^{L\mathcal{B}(4)}_2(Q^2, x_B, x_L, t) \notag \\
&= \frac{4\pi \alpha^2}{x_B Q^4} \left(1-y+\frac{y^2}{2}\right) F^{\mathcal{M}}_2 \left(\frac{x_B}{1-x_L}, Q^2\right) f_{\mathcal{M}/p}(x_L, t),
\label{eq:CS}
\end{align}

where $\mathcal{M} = \pi^+$ or $K^+$, and $F^{L\mathcal{B}(4)}_2(Q^2, x_B, x_L, t)$ denotes the semi-inclusive structure function. In this expression, the leading-baryon structure function $ F_2^{L\mathcal{B}(4)} $ factorizes into the meson structure function $F_2^\mathcal{M}$ and the meson flux surrounding the proton, $f_{\mathcal{M}/p}$.

In the effective field theory, the meson flux is typically evaluated at the pion or kaon pole~\cite{McKenney:2015xis,Goncalves:2015mbf,Xie:2021ypc, Wang:2023thl, Xie:2020uck,Bishari:1972tx,Holtmann:1994rs,Kopeliovich:1996iw},
\begin{equation}
    f_{\mathcal{M}/p}(x_L, t) = \frac{1}{2\pi}\frac{g^2}{4\pi}\left(1-x_L\right)\frac{-t}{(m^2_\mathcal{M} -t)^2} \text{exp}\left(\mathcal{C}R^2_{\mathcal{B}\mathcal{M}}\frac{t-m^2_\mathcal{M}}{1-x_L}\right),
    \label{eq:SF}
\end{equation}
\end{widetext}
where $m_\mathcal{M}$ is the meson mass. For pions and kaons, the relevant coupling constants are $ g = g_{pn\pi} $ and $ g = g_{N\Lambda K} $, with $ g^2_{pn\pi}/4\pi = 13.6$  \cite{Timmermans:1990tz} and $ g^2_{N\Lambda K}/4\pi = 14.7 $ \cite{Xie:2021ypc}, respectively. The interaction radii are $ R_{n\pi} = 0.93 ~\text{GeV}^{-1}$ \cite{Holtmann:1994rs} for the $n-\pi$ system and $ R_{\Lambda K} = 1.0 ~\text{GeV}^{-1}$ \cite{Xie:2021ypc} for the $\Lambda-K$ system. The parameter $\mathcal{C}$ is an auxiliary sign factor introduced to unify the expression: $\mathcal{C} = 1$ for the $n-\pi$ system and $\mathcal{C} = -1$ for the $\Lambda-K$ system. \par

\section{EicC detector}
\label{sec:detector}
EicC is a proposed high-energy nuclear physics facility designed to probe the internal structure of hadrons, with particular emphasis on the sea-quark region, as well as to investigate exotic hadronic states and the properties of nuclear matter \cite{cao2020eicc, Anderle:2021wcy,Chen:2018wyz}. With a tunable center-of-mass energy of 15–20~GeV and a luminosity exceeding $10^{33}$~cm$^{-2}$s$^{-1}$, EicC will enable unprecedented precision in measurements of sea-quark dynamics. The collider is designed to operate by default with a 3.5~GeV electron beam and a 20~GeV proton beam, achieving a luminosity nearly two orders of magnitude higher than that of the former HERA \cite{DeBoer:2016tnn} collider in Germany, thereby providing a wealth of high-precision data on hadron structure. A central detector and a set of forward detectors constitute key components of the EicC experimental setup. The central detector provides large acceptance, with a pseudorapidity coverage from $\eta = -3.5$ to 3.5, enabling comprehensive tracking and particle identification over a wide kinematic region. The forward detectors are optimized for detecting leading baryons and mesons produced at high pseudorapidity, particularly from beam remnants and diffractive processes. Complementary to the U.S. Electron–Ion Collider, which focuses on the gluon-dominated regime at higher energies, EicC will play a crucial role in advancing the next era of precision nuclear physics research.\par

To assess the feasibility of measuring the pion and kaon structure functions at the EicC via the Sullivan process, we performed dedicated Monte Carlo simulations focusing on channels involving pion and kaon structure. The detector response was modeled using the EicC fast simulation framework~\cite{Zhu:2024iwa,Anderle:2023uvi,GEANT4:2002zbu}. Event generation was carried out with the \texttt{Tagged-neutron-DIS}~\cite{Xie:2020uck,tagged-neutron-DIS} and \texttt{Tagged-Lambda-DIS}~\cite{Xie:2021ypc,tagged-Lambda-DIS} generators, both implemented in \texttt{C++} and based on the CERN \texttt{ROOT} framework~\cite{Brun:1997pa}. For the LN-DIS process, the generator incorporates comparisons among the global JAM18 fit~\cite{Barry:2018ort}, available experimental data, and predictions from the dynamical parton model. In particular, the pion PDFs were taken from the \texttt{piIM-Parton} set~\cite{Han:2018wsw,piIMParton}, derived within the dynamical parton distribution framework. For the L$\Lambda$-DIS process, where information on the kaon structure is highly limited and only a few global kaon PDF analyses exist \cite{Barry:2025wjx,Bourrely:2023yzi,Chang:2024rbs,Han:2020vjp}, the generator employs a kaon PDF obtained from a model-dependent analysis constrained solely by the NA3 data \cite{Saclay-CERN-CollegedeFrance-EcolePoly-Orsay:1980fhh}, also within the dynamical parton model framework.

\begin{figure*}
  \centering
  \begin{minipage}{0.49\textwidth}
    \centering
    \includegraphics[width=7.5cm]{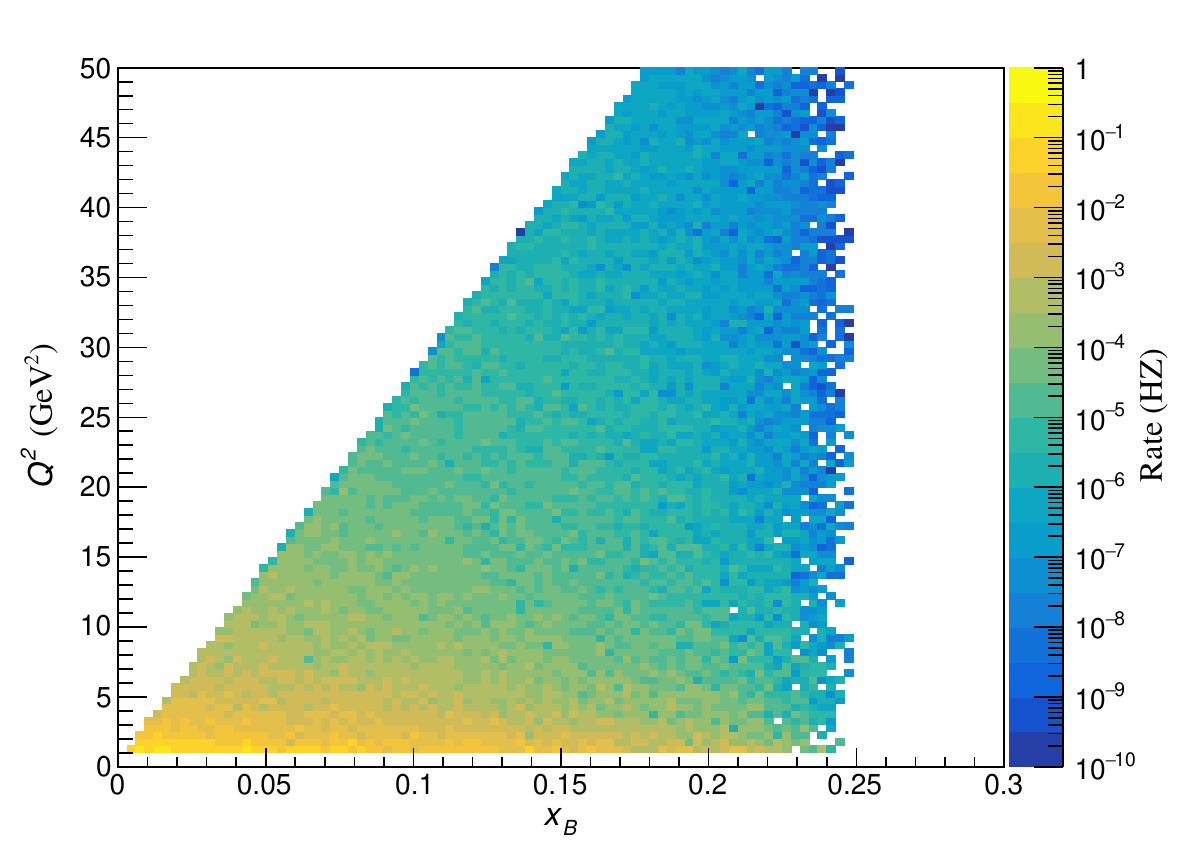}
  \end{minipage}\hfill 
  \begin{minipage}{0.49\textwidth}
    \centering
    \includegraphics[width=7.5cm]{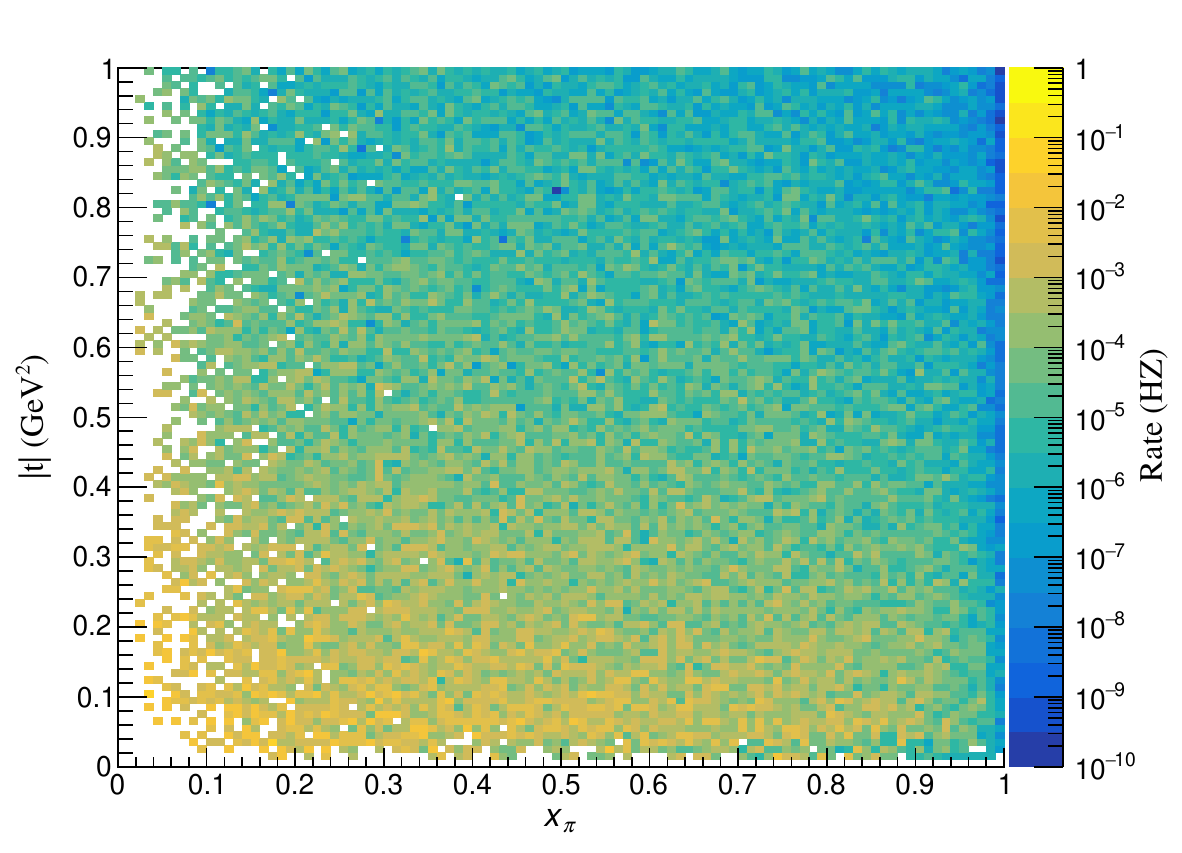}
  \end{minipage}
  \caption{Event rates for the LN-DIS process at the EicC, obtained from Monte Carlo simulation and shown as two-dimensional histograms of invariant kinematic variables. Left panel: distribution in $(x_B, Q^2)$. Right panel: distribution in $(x_\pi, |t|)$. The MC events were generated within the kinematic region $0.01\ \mathrm{GeV}^2<|t|<1\ \mathrm{GeV}^2$, $0.75<x_L<1$, $x_B<1$, $1\ \mathrm{GeV}^2<Q^2<50\ \mathrm{GeV}^2$, and $W^2>4\ \mathrm{GeV}^2$. }
  \label{fig:LNkinematics}
\end{figure*}

\begin{figure*}
  \centering
  \begin{minipage}{0.49\textwidth}
    \centering
    \includegraphics[width=7.5cm]{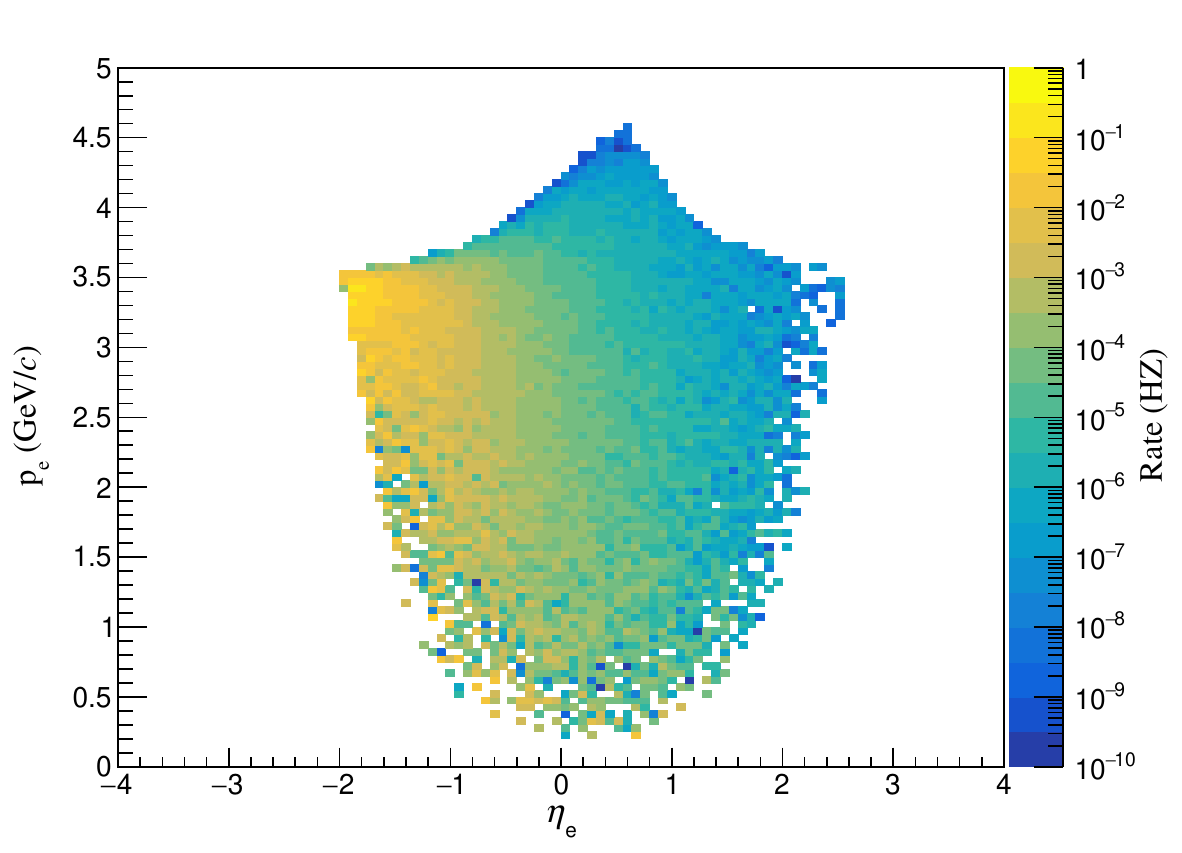}
  \end{minipage}\hfill 
  \begin{minipage}{0.49\textwidth}
    \centering
    \includegraphics[width=7.5cm]{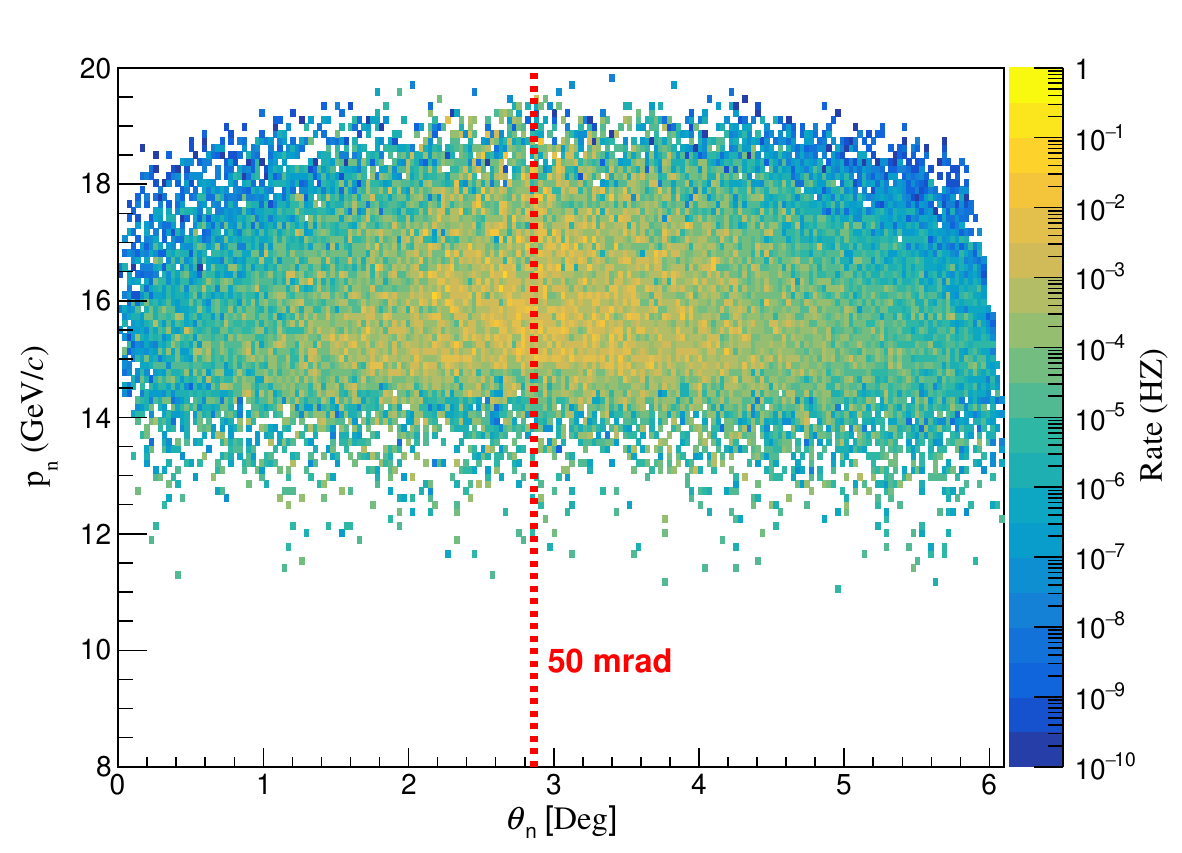}
  \end{minipage}
  \caption{For the Sullivan process $ep \to e n X$, the left panel shows the correlation between the scattered electron momentum and pseudorapidity $\eta$, while the right panel depicts the relationship between the detected neutron momentum and scattering (polar) angle. The distributions are approximately symmetric with respect to the incident proton beam, which is designed to have a crossing angle of 50~mrad (about 2.86°). All cases correspond to the energy setting of a 3.5~GeV electron colliding with a 20~GeV proton. The distributions for the $ep \to e \Lambda X$ process exhibit a similar pattern.}
  \label{fig:LNAngle}
\end{figure*}

\begin{figure*}[htbp]
    \centering
    \begin{adjustbox}{angle=-90}
    \includegraphics[width=15cm, height=16cm, keepaspectratio]{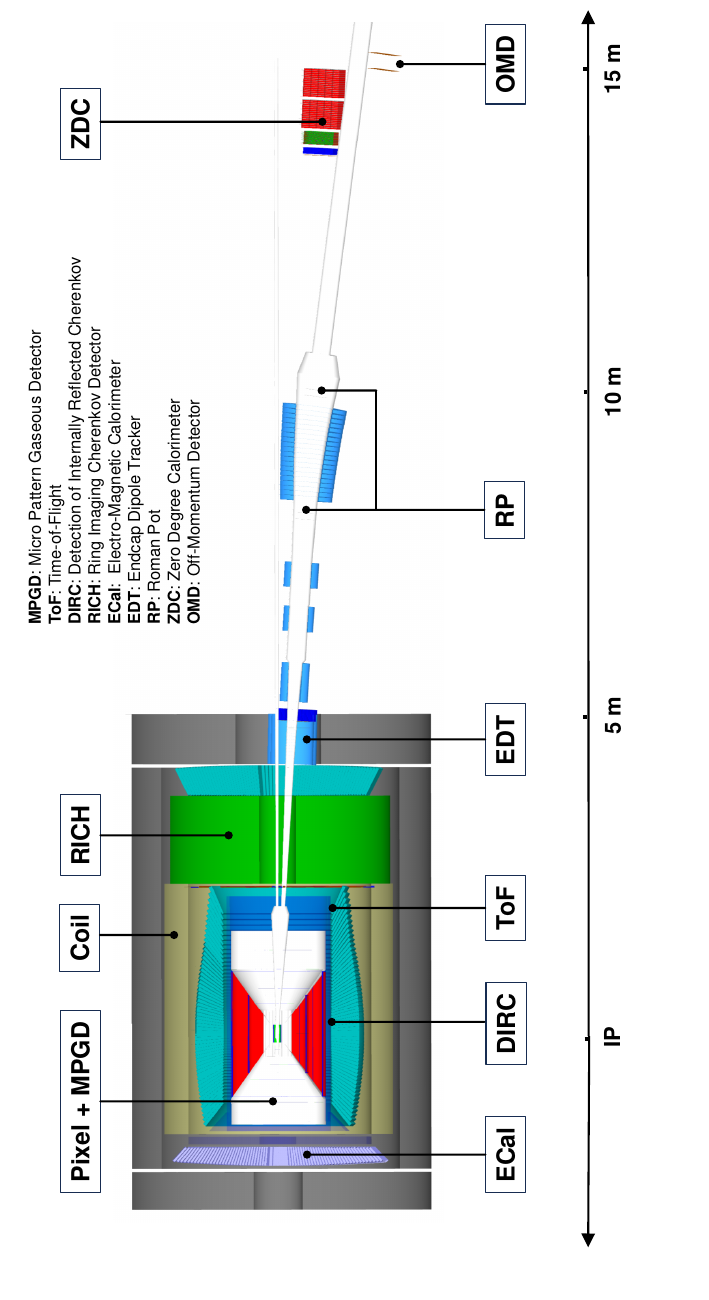}
    \end{adjustbox}
    \caption{Conceptual design of the EicC detector, with the central and ion far-forward detectors included.}
    \label{detector layout}
\end{figure*}
Figure~\ref{fig:LNkinematics} illustrates the event distributions of the generated LN-DIS process within the selected kinematic region. As shown in the left panel, the events populate a broad kinematic phase space, with a higher density observed at relatively small $Q^2$ and $x_B$, where the virtual photon predominantly probes low-momentum partons in the meson cloud. Under the applied selection cuts, $x_B$ extends up to approximately 0.25. In contrast, the right panel displays the event distribution in the $(x_\pi, t)$ plane, covering almost the entire accessible kinematic range. A clear correlation is visible, with the event density gradually decreasing from the small-$x_\pi$, small-$|t|$  region toward the large-$x_\pi$, large-$|t|$region. This behavior reflects the characteristic $t$-dependence of the meson-exchange mechanism inherent to the Sullivan process. Overall, these distributions confirm the expected kinematic features of meson-exchange–induced DIS and demonstrate the feasibility of probing meson structure via leading-baryon tagged measurements at the EicC. In the LN-DIS process, the scattered electron and the produced hadronic system originate from a hard interaction between the virtual photon and the meson cloud surrounding the proton. As shown in Figure~\ref{fig:LNAngle}, the scattered electrons are primarily distributed within the central rapidity region ($|\eta| < 3.5$) with a broad momentum range, where the central detector of the EicC provides precise tracking and particle identification with an efficiency exceeding 95\%. In contrast, the final-state neutron emerges at a very small angle relative to the incident proton beam, typically within a few tens of milliradians, while carrying a large fraction of the proton’s initial momentum. This kinematic feature underscores the need for a dedicated far forward detection system.

As illustrated in Fig.~\ref{detector layout}, the EicC far-forward detector system consists of several sub-detectors, including the Endcap Dipole Tracker (EDT), Roman Pots (RP), Off-Momentum Detector (OMD), and the Zero Degree Calorimeter (ZDC). These detectors operate collaboratively to reconstruct neutrons, photons, and charged hadrons produced at extremely small scattering angles.

The ZDC detects neutrons within approximately $15~\text{mrad}$ of the ion-beam direction and provides sufficient performance for reconstructing the four-momentum transfer $t$. According to simulation studies, the ZDC achieves an energy resolution of
\begin{equation}
    \label{ZDC_E_reso}
    \frac{\sigma_E}{E} \approx \frac{48.7\%}{\sqrt{E~[\text{GeV}]}} + 1.9\% ,
\end{equation}

and a transverse position resolution of
\begin{equation}
    \label{ZDC_P_reso}
    \sigma_x \approx \frac{3~\text{cm}}{\sqrt{E~[\text{GeV}]}} ,
\end{equation}

corresponding to an angular resolution of approximately $2.1~\text{mrad}/\sqrt{E~[\text{GeV}]}$ at a distance of 14 m from the IP. This performance enables a typical $t$ resolution of about $0.03~\text{GeV}^2$.

The EDT, RP, and OMD jointly reconstruct the charged decay products of the $\Lambda$ baryon produced in the L$\Lambda$-DIS process.  
The $\Lambda$ decays predominantly via the charged channel ($\Lambda\to p\pi^-$, BR~$\sim64\%$) and the neutral channel ($\Lambda\to n\pi^0\to n\gamma\gamma$, BR~$\sim36\%$).  
Reconstructing this topology requires the combined acceptance of all forward detectors.

For charged decay channel, the proton is reconstructed using the combined tracking information from the EDT, RP, and OMD.  
The EDT provides tracking in $15~\text{mrad} < \theta < 60~\text{mrad}$ with a momentum resolution of
\begin{equation}
    \frac{\sigma_p}{p}\approx0.6\% ,
\end{equation}
and a polar-angle resolution better than $3~\text{mrad}$.  
The RP extends charged-particle tracking down to $5~\text{mrad}$ and achieves an improved momentum resolution of
\begin{equation}
    \frac{\sigma_p}{p}\approx0.2\% .
\end{equation}
The OMD captures low-momentum decay protons with $0.4 < x_L < 0.75$, requiring a spatial resolution of order $100~\mu\text{m}$, achievable with Monolithic Active Pixel Sensors (MAPS) or Micro-Pattern Gaseous Detectors (MPGDs).  
The $\pi^-$ from the decay is accepted and reconstructed solely by the EDT tracking system.

For neutral decay channel, the neutron is tagged by the ZDC, while the two photons from the $\pi^0$ decay are jointly reconstructed using the ZDC, the EDT electromagnetic calorimeter (EDT-ECal), and the hadron-endcap ECal of the central detector.

The capability of reconstructing $\Lambda$ baryons over a wide kinematic range is particularly relevant for studies of the kaon structure via the Sullivan process. In such measurements, the final-state $\Lambda$ carries critical information about the kaon, and efficient reconstruction of its decay products directly impacts the precision and reach of the structure-function extraction. Therefore, the performance of the far-forward detector system—both in charged and neutral decay channels—is a key factor in determining the sensitivity of kaon-structure measurements.

\begin{figure*}[htbp]
    \centering
    \includegraphics[width=0.6\textwidth]{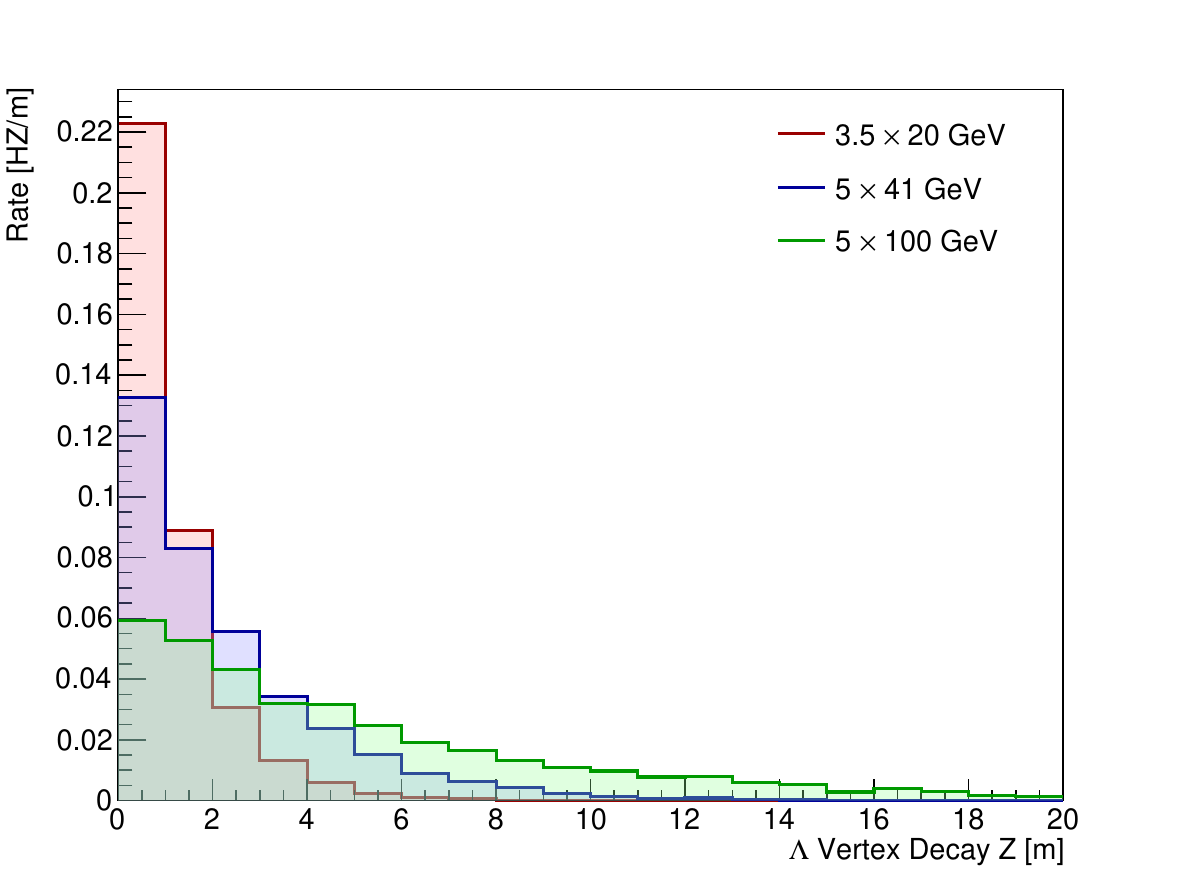}
    \caption{Distribution of the $\Lambda$ decay vertex along the $z$ direction for final-state $\Lambda$ at different collision energies. }
    \label{fig:LambdaDecaVz}
\end{figure*}

With this capability in place, the EicC offers unique advantages for studying the kaon structure. Since the kaon structure is accessed via the Sullivan process, the reconstruction of the $\Lambda$ baryon in the final state is essential. The $\Lambda$ has a relatively long lifetime of $\tau = (2.617 \pm 0.010) \times 10^{-10}~\text{s}$~\cite{ParticleDataGroup:2024cfk}, meaning that at high beam energies its decay length can easily exceed several tens of meters. Figure~\ref{fig:LambdaDecaVz} shows the distribution of the $\Lambda$ decay vertex along the flight direction at different collision energies. For the typical EicC configuration of $3.5 \times 20~\text{GeV}$, more than 95\% of $\Lambda$ baryons decay within 5~m of the interaction point—well within the coverage of the forward detector system. In contrast, for the lower-energy EIC configuration of $5 \times 41~\text{GeV}$, only about 80\% of $\Lambda$ baryons decay within 5~m, and this fraction drops below 60\% for higher energies such as $5 \times 100~\text{GeV}$. Consequently, under comparable luminosity conditions, the EicC detector is expected to have a higher detection efficiency for kaon-structure measurements. Moreover, the larger crossing angle allows for a more spacious detector design, which further enhances the overall detection efficiency.

\begin{figure*}
  \centering

  \begin{minipage}{0.49\textwidth}
    \centering
    \includegraphics[width=8cm]{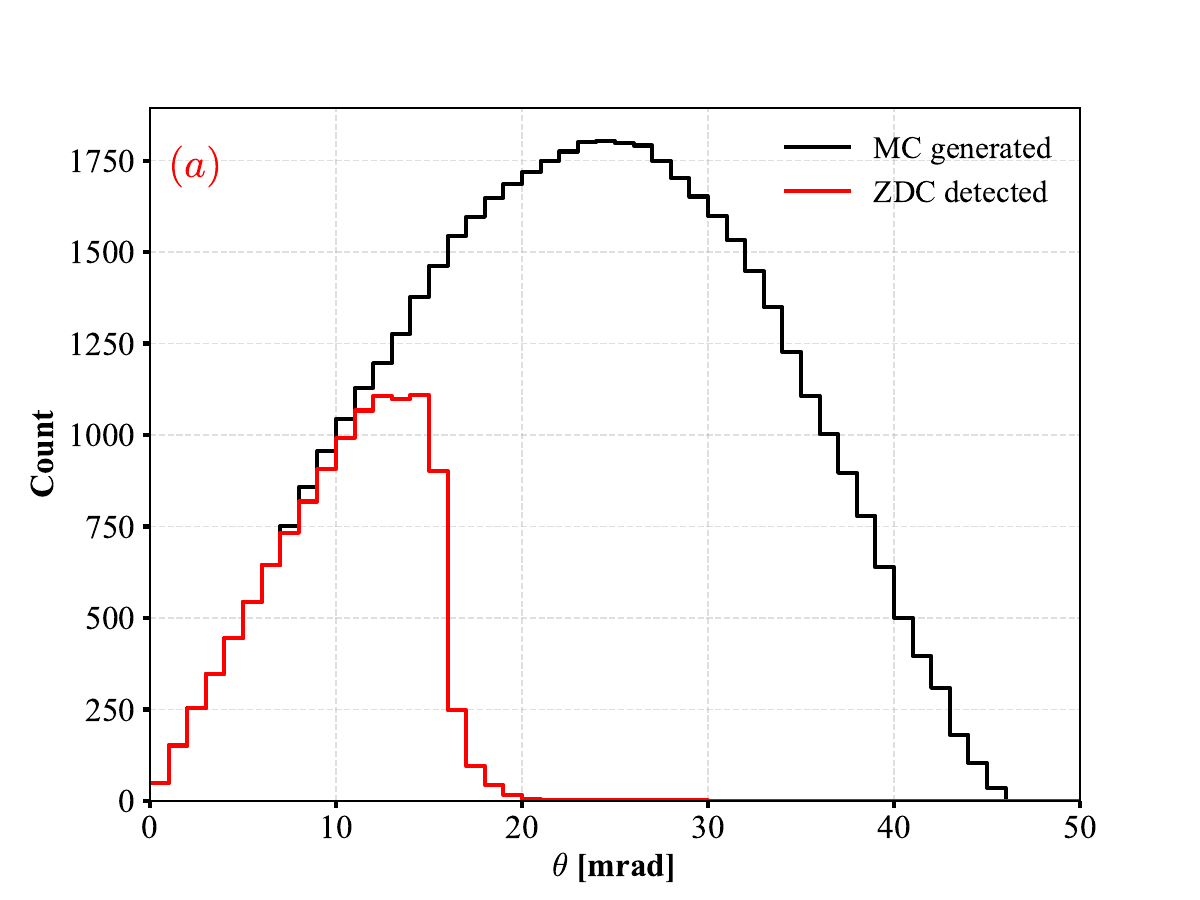}
  \end{minipage}\hfill 
  \begin{minipage}{0.49\textwidth}
    \centering
    \includegraphics[width=8cm]{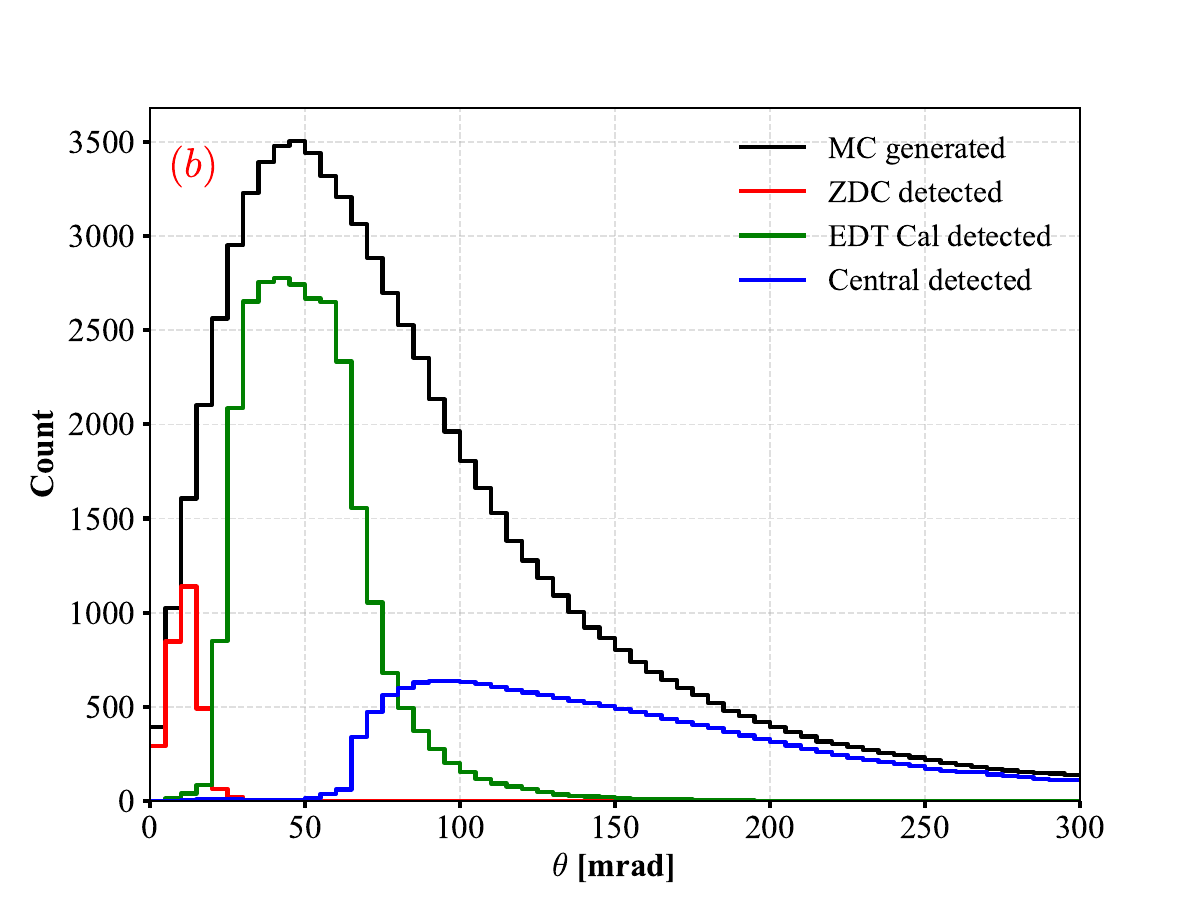}
  \end{minipage}
  \begin{minipage}{0.49\textwidth}
    \centering
    \includegraphics[width=8cm]{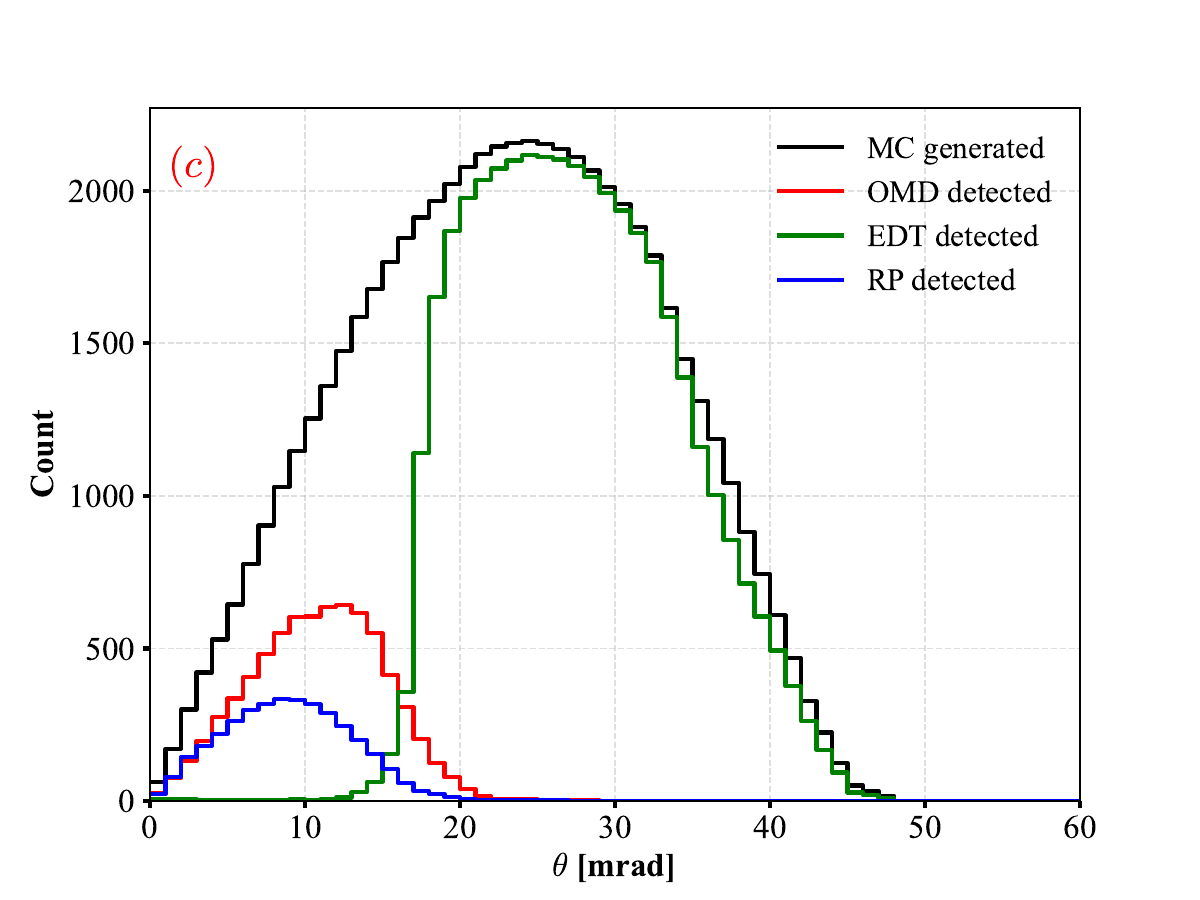}
  \end{minipage}\hfill 
  \begin{minipage}{0.49\textwidth}
    \centering
    \includegraphics[width=8cm]{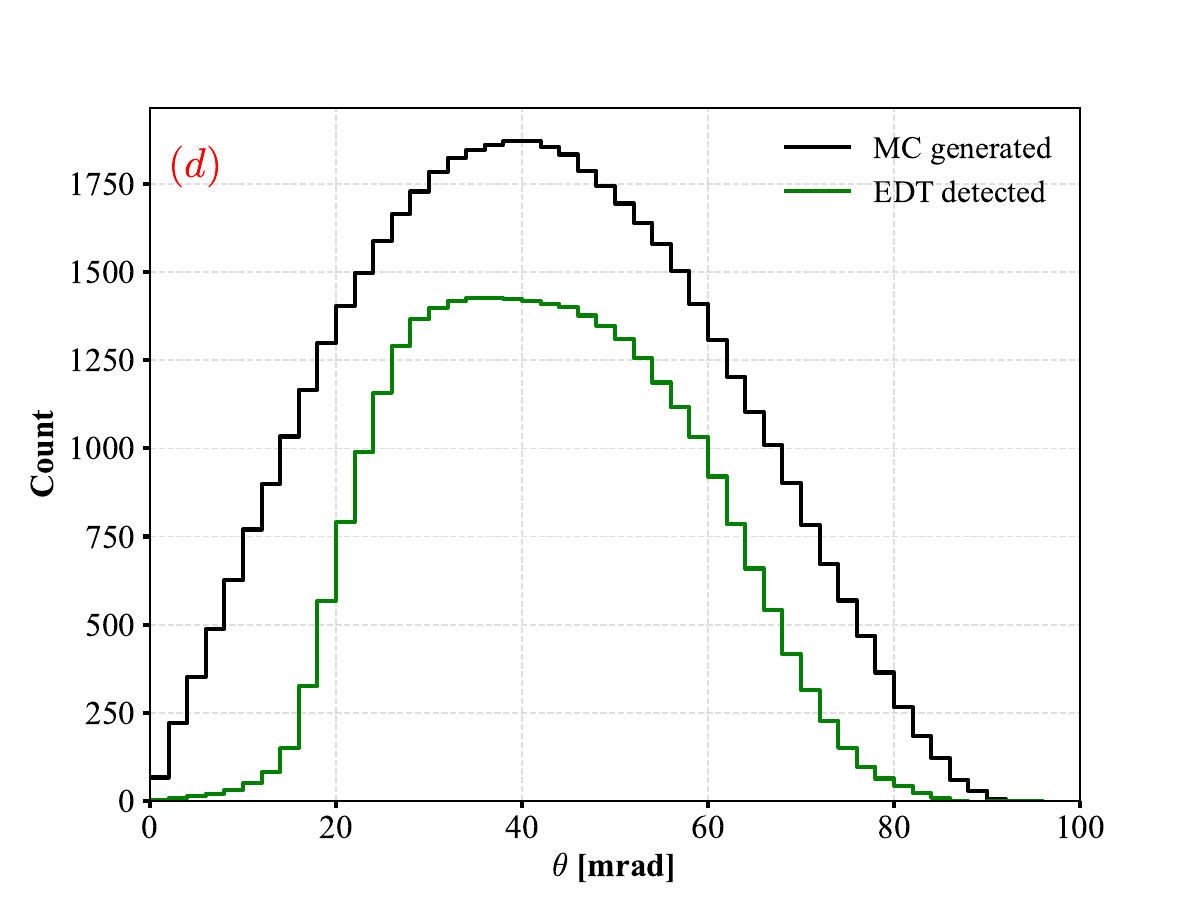}
  \end{minipage}

  \caption{Angular distributions and detector acceptances for neutrons $(a)$ and photons $(b)$ from the neutral $\Lambda$ decay mode, and for decay protons $(c)$ and $\pi^-$ mesons $(d)$ from the charged $\Lambda$ decay mode, for a 20 GeV proton beam on a 3.5 GeV electron beam.}
  \label{Angular_Dis}
\end{figure*}

Building on these kinematic and detector advantages, the reconstruction performance of $\Lambda$ decays at EicC is illustrated in Figure~\ref{Angular_Dis}, which shows the angular distributions of the $\Lambda$ decay products—$(a)$ neutron and $(b)$ photon from the neutral decay channel, and $(c)$ proton and $(d)$ $\pi^-$ from the charged decay channel—together with their corresponding acceptance in the far forward detector system. Building upon these reconstruction capabilities, the kaon structure function can be measured following a strategy analogous to that used for the pion. By focusing on the charged decay channel, which dominates the statistics and offers higher detection efficiency, the impact of detector-resolution effects is mitigated compared with the pion measurement, as both the final-state proton and pion can be reconstructed with high precision. Overall, the full far-forward detector system yields an approximate $\Lambda$ detection efficiency of about $40\%$, with the charged decay channel contributing roughly $55\%$ and the neutral channel about $15\%$.

\section{Structure Function}

\begin{figure*}
  \centering
  \begin{minipage}{0.5\textwidth}
    \centering
    \includegraphics[width=8cm]{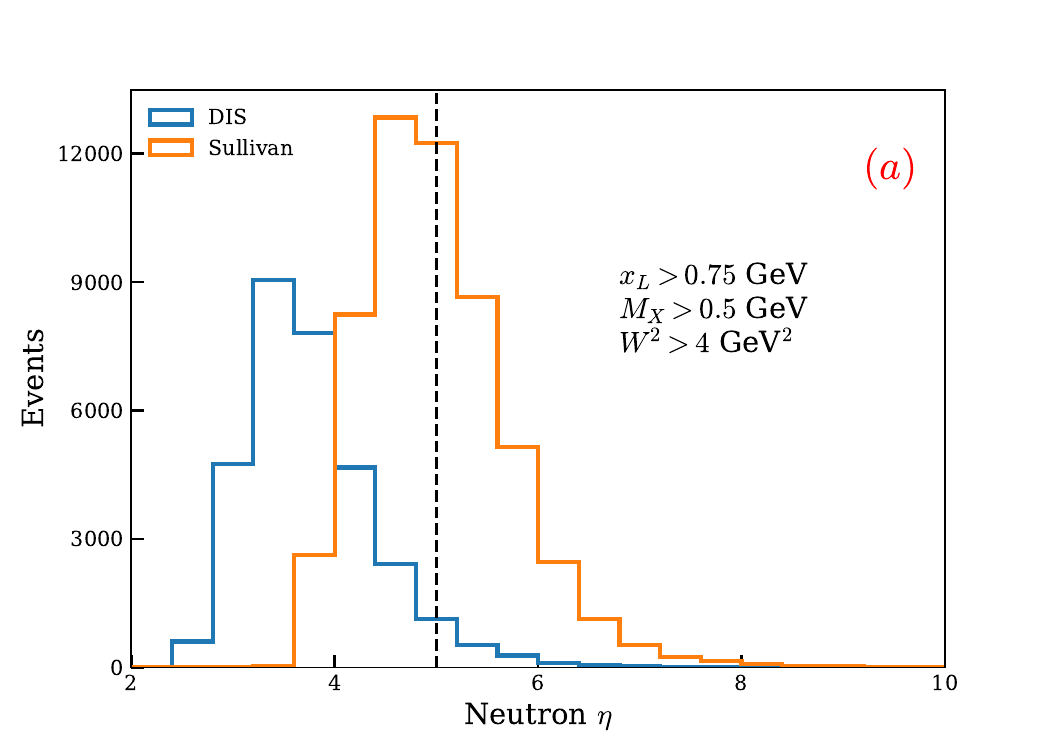}
  \end{minipage}\hfill 
  \begin{minipage}{0.5\textwidth}
    \centering
    \includegraphics[width=8cm]{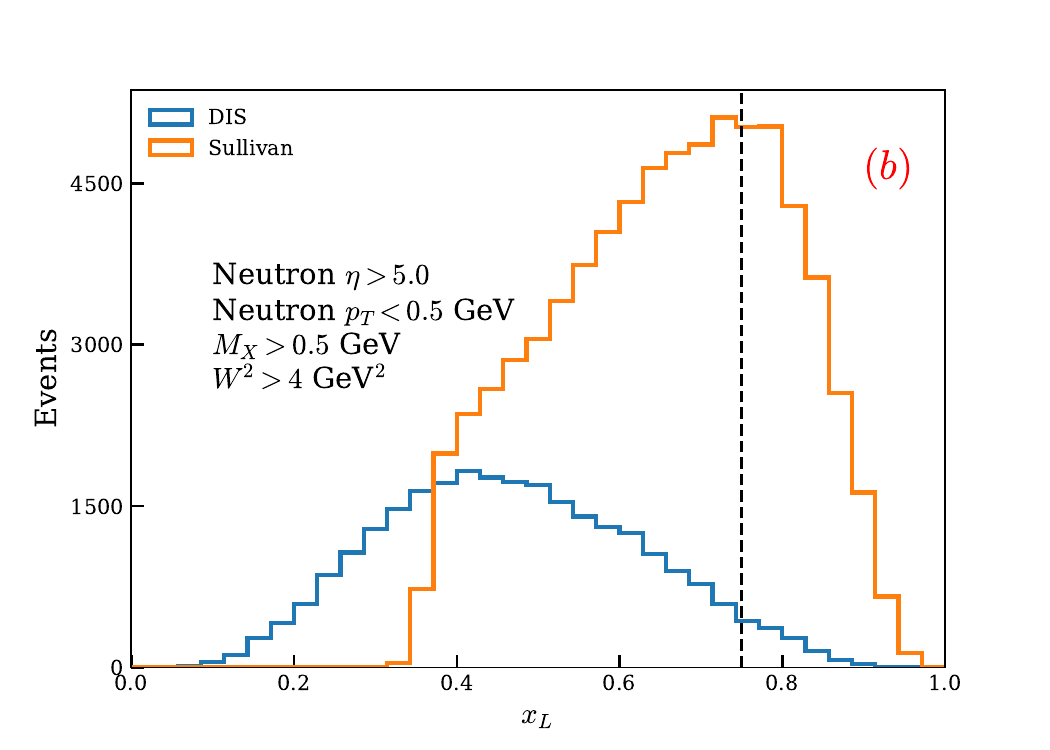}
  \end{minipage}
  \caption{Kinematic distributions of the final-state neutron in DIS (blue) and Sullivan (orange) events.
$(a)$: Pseudorapidity ($\eta$) distribution of the neutron after applying $x_L > 0.75$, $M_X > 0.5$~GeV, and $W^2 > 4$~GeV$^2$. The dashed line at $\eta = 5$ indicates the selection cut used to achieve high Sullivan purity.
$(b)$: Distribution of $x_L$ for the neutron after applying $\eta > 5$, $p_T < 0.3$~GeV, $M_X > 0.5$~GeV, and $W^2 > 4$~GeV$^2$. The dashed line at $x_L = 0.75$ indicates the final selection threshold used to achieve high Sullivan purity.}
  \label{fig:dis_sullivan_eta_xL}
\end{figure*}

To reliably extract the meson structure functions via the Sullivan process, it is essential to isolate events of insterest with high purity. Figure~\ref{fig:dis_sullivan_eta_xL} illustrates the key kinematic distributions used to separate Sullivan events from generic DIS backgrounds, with the DIS sample generated using the \texttt{PYTHIA 8} event generator \cite{Bierlich:2022pfr}. As shown in Fig.~\ref{fig:dis_sullivan_eta_xL}(a), after applying the kinematic requirements $x_L > 0.75$, $M_X > 0.5$~GeV, and $W^2 > 4$~GeV$^2$, the pseudorapidity distribution of the final-state neutron exhibits a pronounced peak in the far-forward region, showing a clear dominance of events at $\eta_n > 4$, which enables efficient background suppression. The ZDC covers the region $\eta_n > 5$, which naturally helps to suppress background. Figure~\ref{fig:dis_sullivan_eta_xL}$(b)$ shows the $x_L$ distributions of DIS and Sullivan events after applying the cuts $\eta_n > 5$, $p_T^n < 0.3$~GeV, $M_X > 0.5$~GeV, and $W^2 > 4$~GeV$^2$. The $x_L$ variable is a powerful discriminator for the two different processes, with the Sullivan events peak strongly at large $x_L$, whereas DIS backgrounds fall off at high $x_L$. A final requirement of $x_L > 0.75$ and $\eta_n > 5$, indicated by the vertical dashed lines, yields a Sullivan-event purity exceeding 90\%. In addition, conventional cuts of $0.01 < |t| < 1~\text{GeV}^2$ and $x_B < 1$ were applied to ensure that the simulated data follow physically realistic distributions.

The extraction of the meson structure functions is performed in bins of $(x_{\mathcal{M}}, Q^{2})$. 
The event generator provides, for each generated event, the differential cross section and the corresponding event rate. 
When filling the histograms, each event is weighted by the product of its rate and the assumed integrated luminosity, such that the resulting distributions represent the expected yield under realistic running conditions.

For each $(x_{\mathcal{M}}, Q^{2})$ bin, the differential yield as a function of the squared momentum transfer $t$ of the forward baryon is constructed. 
The relation between the four-fold differential cross section and the meson structure function is given by Eq.~\ref{eq:CS}, which factorizes the measured cross section into the structure function $F_2^{\mathcal{M}}$ and the meson flux factor $f_{\mathcal{M}/p}(x_L, t)$. 
In the extraction, the flux $f_{\mathcal{M}/p}(x_L, t)$ is fixed using a chosen flux model, while $F_2^{\mathcal{M}}$ is determined by fitting the $t$ distribution with Eq.~\ref{eq:CS}. The statistical uncertainty of $F_2^{\mathcal{M}}$ is taken from the covariance matrix of the fit.

The systematic uncertainty assessment has been performed with particular emphasis on detector-related effects. These studies were first carried out for the pion measurement, followed by a parallel evaluation for the kaon measurement.

The important source of systematic uncertainty arises from detector resolution effects. These were implemented by applying Gaussian smearing to the true four-momentum vectors of both the scattered electron and the forward neutron. For the scattered electron, the momentum, polar angle, and azimuthal angle were independently smeared according to the parameterized resolutions obtained from the EicC fast simulation. For the forward neutron detected in the ZDC, the energy and impact position on the detector plane were smeared using the energy-dependent resolutions given by the position resolution in Eq.~\ref{ZDC_P_reso} and the energy resolution in Eq.~\ref{ZDC_E_reso}. The resulting smeared four-momentum vectors were propagated through the reconstruction chain to incorporate realistic detector effects.

To estimate the uncertainty associated with these resolutions, the energy and position smearing parameters were varied by $\pm 10\%$, and the induced variations in the reconstructed distributions were taken as the corresponding systematic uncertainty. As an additional cross-check, both smeared and unsmeared distributions were fitted using Eq.~\ref{eq:CS}. The detector-resolution uncertainty contributes approximately 10\% to the pion measurement.

Beyond detector resolution, several additional sources of systematic uncertainty were evaluated. A 3\% uncertainty arises from luminosity normalization, reflecting the precision of the integrated luminosity determination. Acceptance corrections introduce another 8\% uncertainty due to potential mismodeling of the detector acceptance across the relevant kinematic range. The ratio method \cite{ZEUS:2002gig}, which extracts the pion structure function by taking the ratio of tagged to inclusive DIS yields to cancel common systematic effects, introduces an additional 2\% uncertainty associated with the proton structure function $F_2^p$.  These contributions together form the complete systematic uncertainty budget for the pion measurement.

In the analysis of the kaon structure function, we use only the charged decay channel for now for simplicity. The associated systematic uncertainty study is performed following the same methodology as in the pion case. Detector-resolution effects were evaluated by applying the identical smearing procedure to the kaon sample. However, because the reconstructed final state predominantly comes from the decay $\Lambda \to p\pi^{-}$, whose charged decay products are efficiently measured by the tracking system, the overall momentum resolution is significantly improved. Consequently, the detector-resolution uncertainty for the kaon measurement is reduced to approximately 5\%.

In addition, the background process $e\,p \to e'\,\Sigma^{0} X$, $\Sigma^{0} \to \Lambda \gamma$ introduces a non-negligible systematic uncertainty to the kaon structure--function measurement. Since both the signal process $e p \to e \Lambda X$ and the background channel produce a $\Lambda$ baryon in the final state, the contamination from $\Sigma^{0}$ production must be taken into account. To estimate this effect, we compared the hadronic couplings $g_{K\Lambda N}$ and $g_{K\Sigma N}$ using values reported in various theoretical studies \cite{Aliev:1999mj,Adelseck:1990ch,Williams:1991tw,Williams:1992tp,Mart:1995wu,David:1995pi,General:2003sm,Guidal:1997hy,Lee:1999kd, Martin:1980qe, Gobbi:1992tk, Choe:1995yb,General:2003sm,Chiang:2001pw}, which suggests that the relative size of the $\Sigma^{0}$ contribution could be at the level of approximately $10\%$. However, the charged decay mode  $\Sigma^{0}\to\Lambda\gamma$ can be partially vetoed when the decay photon from $\Sigma^0$ has sufficient energy. Taking this into consideration, we adopt a conservative estimate of a residual $\sim 5\%$ systematic uncertainty associated with the $\Sigma^{0}$ background.

All remaining sources of systematic uncertainty—including luminosity normalization, acceptance effects, and uncertainties from the ratio method—are treated using the same procedures as in the pion analysis, ensuring full consistency across both measurements.

\begin{figure*}
  \centering
  \begin{minipage}{0.49\textwidth}
    \centering
    \includegraphics[width=7.5cm,trim={0 0 0 0},clip,keepaspectratio]{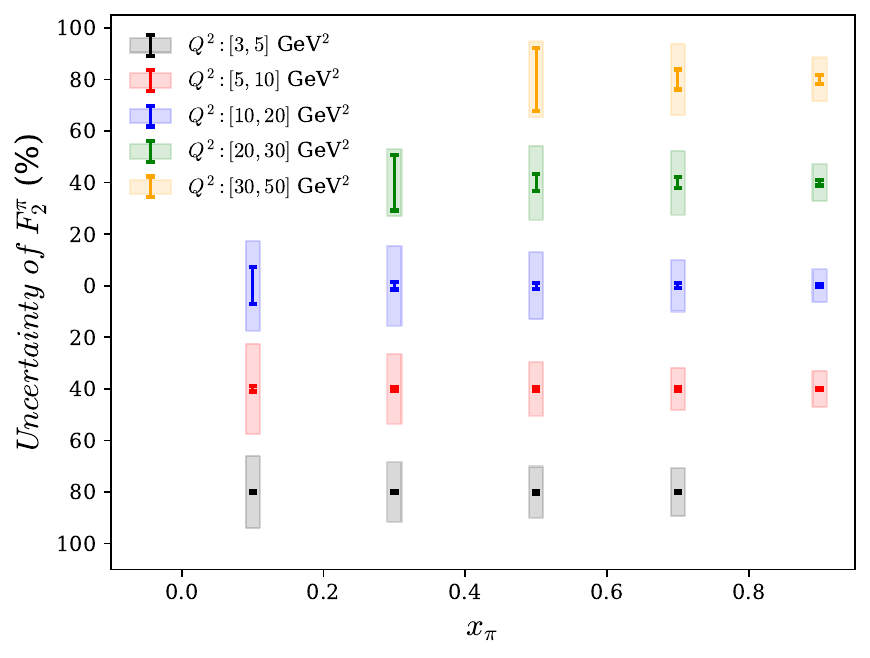}
  \end{minipage}\hfill 
  \begin{minipage}{0.49\textwidth}
    \centering
    \includegraphics[width=7.5cm,trim={0 0 0 5},clip,keepaspectratio]{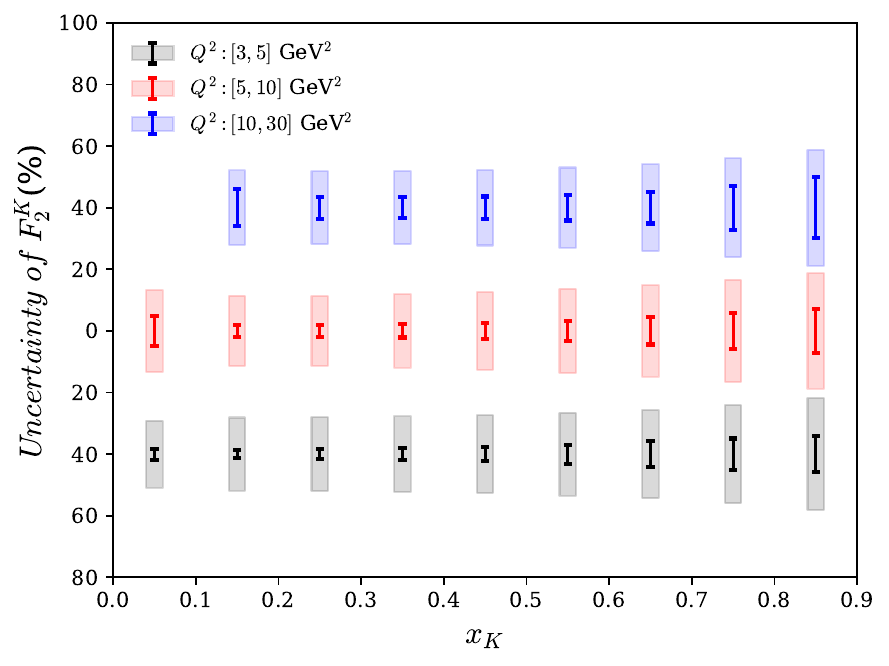}
  \end{minipage}
  \caption{\textbf{Left panel}: Projected pion structure function $F_2^{\pi}$ at EicC, showing statistical uncertainties (error bars) and systematic uncertainties (color bands, excluding the flux factor) for different $x_{\pi}$ and $Q^2$ bins. \textbf{Right panel}: Projected kaon structure function $F_2^{K}$ at EicC, showing similar analysis results. Both cases assume an integrated luminosity of $50~\text{fb}^{-1}$ and an energy setting of $3.5 \times 20~\text{GeV}^2$.
}
  \label{structurefunctionUnc}
\end{figure*}

Based on all the sources of uncertainty discussed above, Fig.~\ref{structurefunctionUnc} presents the systematic and statistical uncertainties for the pion (left panel) and kaon (right panel) structure--function extractions across the relevant kinematic ranges. For the pion measurement, the accessible kinematics span $Q^{2}=3\text{--}50~\text{GeV}^{2}$ and $x_{\pi}=0\text{--}1$. For $Q^{2}>5~\text{GeV}^{2}$, the statistical uncertainty in each analysis bin falls below $5\%$, reaching the precision necessary for detailed studies of the pion structure function. In this channel, the ZDC dominates the systematic uncertainty, with its energy and position resolutions discussed in Sec.~\ref{sec:detector}, and the reconstruction of the squared momentum transfer achieving a resolution of approximately $\Delta t \sim 0.03~\text{GeV}^{2}$. For the kaon measurement (right panel), the analysis employs the charged decay channel $\Lambda \to p\pi^{-}$, achieving $Q^{2}$ coverage up to $30~\text{GeV}^{2}$. Thanks to the charged decay tracks, the momentum transfer $t$ of the $\Lambda$ can be reconstructed with a resolution of about $\Delta t \sim 0.0075~\text{GeV}^{2}$, and the overall reconstruction efficiency of the charged decay exceeds $50\%$. The statistical uncertainty remains below $8\%$ for all bins, enabling precise studies of the kaon structure function. Overall, these results demonstrate that EicC offers unique advantages for kaon structure--function measurements through the Sullivan process.

\begin{figure*}[htbp]
    \centering
    \begin{adjustbox}{angle=0}
    \includegraphics[width=10cm, keepaspectratio]{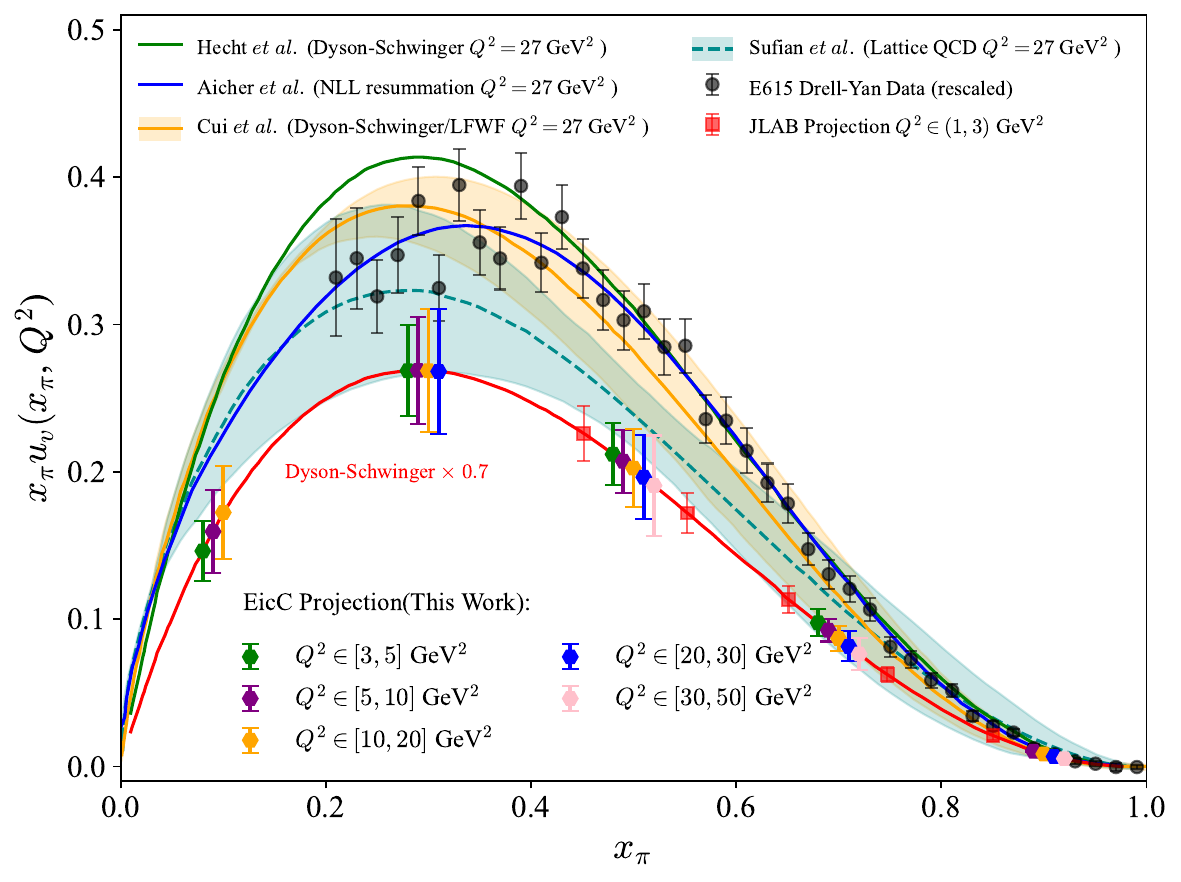}
    \end{adjustbox}
    \caption{Projected measurements of the pion valence up-quark distribution $x_\pi u_v^\pi(x_\pi)$ are shown for several $Q^2$ bins. The EicC projections are represented by colored points with error bars, with the error bars indicating the combined statistical and estimated systematic uncertainties. For comparison, the Drell--Yan E615 data (black circles)~\cite{E615:1989bda}, rescaled following the reanalysis of Refs.~\cite{Aicher:2010cb, Chang:2014lva}, are shown. The JLab projection results at $Q^2 = 1$--$3~\mathrm{GeV}^2$ (red squares)~\cite{Adikaram2015TDIS} are also included for reference. In addition, several representative theoretical predictions at $Q^2 = 27~\mathrm{GeV}^2$ are displayed, including Dyson--Schwinger calculations (green and orange bands/curves)~\cite{Hecht:2000xa, Cui:2020tdf, Nguyen:2011jy}, the NLO reanalysis of Drell--Yan data with soft-gluon resummation (blue solid line)~\cite{Aicher:2010cb, Nguyen:2011jy}, and lattice QCD calculations (teal dashed line and band)~\cite{Sufian:2019bol}. 
    }
    \label{fig:F2pi_comparison}
\end{figure*}

In Fig.~\ref{fig:F2pi_comparison}, we present the projected EicC measurements of the pion valence up-quark distribution, $x_\pi u_v^\pi(x_\pi)$, for several $Q^2$ bins. The EicC projections are represented by colored points, with error bars indicating the combined statistical and estimated systematic uncertainties. For comparison, the Drell--Yan E615 data (black circles)~\cite{E615:1989bda}, rescaled following the reanalysis of Refs.~\cite{Aicher:2010cb,Chang:2014lva}, are shown.  Earlier Drell--Yan measurements from the NA3 and NA10 experiments~\cite{NA3:1983ejh,NA10:1985puz} provide complementary constraints in similar kinematic regions, although their precision and theoretical treatment differ from the reanalyzed E615 data and are therefore not shown here for clarity.  The JLab projection results at $Q^2 = 1$--$3~\mathrm{GeV}^2$ (red squares)~\cite{Adikaram2015TDIS} are also included for reference. All reference curves and uncertainty bands from theoretical or phenomenological analyses are shown at a scale of $Q^2 = 27~\mathrm{GeV}^2$, which matches the scale of the reanalyzed E615 data. These include Dyson--Schwinger calculations by Hecht \textit{et al.}~\cite{Hecht:2000xa} and Cui \textit{et al.}~\cite{Cui:2020tdf}, the NLO reanalysis of Drell--Yan data with soft-gluon resummation by Aicher \textit{et al.}~\cite{Aicher:2010cb}, and lattice QCD calculations by Sufian \textit{et al.}~\cite{Sufian:2019bol}. Several of these results rely heavily on Drell--Yan data in the large-$x_\pi$ region for their phenomenological constraints. In addition, recent global QCD analyses of pion parton distribution functions have been reported by Barry \textit{et al.}~\cite{Barry:2021osv}, and further studies of the E615 Drell--Yan data have been performed by Cui \textit{et al.}~\cite{Cui:2021mom}. These works provide a useful reference for current theoretical and phenomenological studies of pion and kaon parton distributions. The projected EicC measurements, shown across multiple $Q^2$ bins, offer complementary constraints and will help refine our understanding of pion and kaon structure.

To date, this kinematic region remains largely unexplored by Sullivan-process experiments. The EicC, however, provides remarkable sensitivity to the intermediate-to-large $x_\pi$ region in Sullivan measurements, thereby offering crucial new information on the pion structure in the nonperturbative valence regime. With the projected precision demonstrated here, the EicC data can be meaningfully combined with existing and future Drell--Yan measurements. Compared to the current world data, the EicC extends the accessible $Q^2$ coverage by a factor of 2--3 beyond JLab-12~GeV, helping to bridge the gap between fixed-target and collider-era experiments. Since the Drell--Yan process is independent of the meson flux, the extracted structure functions are free from flux-model uncertainties. Therefore, combining Drell--Yan measurements with EicC Sullivan data in the overlapping kinematic region may substantially reduce model-dependent systematic uncertainties, particularly those associated with the pion flux. In practice, the pion flux uncertainty is typically estimated to be at the level of $\sim 10$--$25\%$~\cite{Arrington:2021biu, ZEUS:2002gig, H1:2010hym}. A combined analysis of EicC data with pion Drell--Yan measurements in the same kinematic regime, such as those from AMBER and Fermilab~\cite{NA3:1983ejh, NA10:1985ibr}, is expected to significantly constrain and reduce this source of uncertainty.

While the pion structure function has been relatively well constrained by previous measurements, experimental information on the kaon structure function remains very limited, with only scarce data from Drell-Yan analyses available to constrain its parton distribution \cite{Saclay-CERN-CollegedeFrance-EcolePoly-Orsay:1980fhh}. In contrast, significant theoretical progress has been made in recent years \cite{Holt:2010vj}, including Dyson--Schwinger studies~\cite{Cui:2020tdf,Nguyen:2011jy}, lattice QCD calculations~\cite{LatticeParton:2022zqc, Barry:2025wjx}, and global QCD analyses~\cite{Xu:2024nzp} that incorporate both experimental and lattice constraints. These studies reveal characteristic SU(3) flavor-symmetry breaking effects in the kaon, such as a softer light-quark distribution and a more asymmetric momentum sharing between the valence quarks compared to the pion~\cite{Nguyen:2011jy,LatticeParton:2022zqc, Barry:2025wjx}. Nevertheless, the limited experimental information, particularly in the large-$x$ region, makes it difficult to achieve a definitive determination of the kaon structure. Therefore, precise measurements of the kaon structure function at the EicC will be particularly valuable, providing critical constraints on its partonic structure and enabling stringent tests of nonperturbative QCD dynamics.

\section{Summary and Outlook}

This study investigates the feasibility and expected precision of measuring the meson structure functions $F_2^\pi$ and $F_2^K$ at the proposed EicC, based on the Sullivan process. By selecting final-state forward baryons based on their $x_L$ distributions in the far forward region, high-purity Sullivan events can be identified, with the deep inelastic scattering background effectively suppressed. Within each $(x_M, Q^2)$ bin, the meson structure functions are extracted by fitting the differential yields as a function of the baryon momentum transfer $t$. Detector resolution effects are simulated by applying Gaussian smearing to the four-momenta of the electron and forward baryon, while systematic uncertainties are evaluated through a combination of variations in resolution, acceptance corrections, and other methods. Under the nominal EicC running conditions with an integrated luminosity of 50~fb$^{-1}$, the statistical precision of $F_2^\pi$ reaches better than 5\% in the key kinematic region ($Q^2>5~\mathrm{GeV^2}$), while the overall statistical uncertainty of $F_2^K$ is controlled within 8\%. For $F_2^\pi$, detector-related effects—such as the forward baryon reconstruction efficiency and $t$ resolution—constitute the dominant sources of systematic uncertainty, contributing roughly 10\%. In contrast, for $F_2^K$, these detector effects are mitigated by the presence of charged decay channels, reducing their impact to approximately 5\%.

At the results level, EicC measurements of $F_2^\pi$ exhibit a pronounced advantage in the medium-to-large $x_\pi$ region and can provide complementary constraints to existing Drell--Yan data, thereby further reducing the theoretical uncertainties associated with meson flux factors. In contrast, experimental information on the kaon structure function remains extremely limited; what is exciting is that EicC will provide a particularly valuable opportunity for systematic, high-precision measurements of $F_2^K$ across a broad kinematic range. Leveraging the efficient reconstruction of $\Lambda \to p \pi^-$ decays, superior resolution in momentum transfer can be achieved for the kaon measurement, enhancing the overall feasibility of measuring $F_2^K$.

Our study suggests that, even with the present detector performance assumptions, EicC will serve as an ideal facility for probing the internal structure of pions and kaons. The detector design is still under ongoing optimization. In particular, with more advanced detector technology, the energy resolution of ZDC is expected to reach approximately $35\%/\sqrt {E}$, which will further reduce the systematic uncertainties in the extraction of $F_2^\pi$. For kaon measurements, although the neutral decay channel of the $\Lambda$ has a relatively small branching ratio, it is in principle feasible to reconstruct its decay topology using the forward detector system. At the current stage, however, the main limitation arises from the insufficient size of the ZDC, which leads to a low neutron acceptance, as illustrated in Fig.~\ref{Angular_Dis}$(a)$. To address this issue, one possible improvement is to install an additional hadronic calorimeter downstream of the EDT–ECal system to enhance the acceptance for neutral particles. Such an upgrade would require coordinated design with the beamline. In addition, given the different PID requirements, the neutral decay channel would help constrain the systematic uncertainties in $F_2^K$ due to PID, thereby enhancing the robustness of the meson-structure studies at EicC.

In summary, EicC will provide essential and high-precision experimental input for understanding the quark structure of light mesons under strong interactions, offering a valuable testing ground for non-perturbative QCD.

\section{Acknowledgments}
We would like to express our heartfelt gratitude to Craig D. Roberts, Tanja Horn, Garth M. Huber, Jixie Zhang, Richard Trotta and Avnish Singh for their invaluable theoretical guidance and technical support throughout this work. We also sincerely thank Jianing Dong for ensuring the excellent performance and reliability of the computer facility at Shandong University, which made our computations possible.

This work is supported by the National Natural Science Foundation of China under Grant Nos.~12305144 and 12375139, the Shandong Province Natural Science Foundation under Grant Nos.~2023HWYQ-010 and ZR2022QA032, and the National Key Research and Development Project of China under Contract No.~2023YFA1606800.

\bibliographystyle{apsrev4-2}
\bibliography{reference}
\end{document}